\documentclass{aastex62}

\usepackage{amsmath}
\usepackage{hyperref}
\usepackage{rotating}
\usepackage{color}

\newcommand{\fermi}{\emph{Fermi}}
\newcommand{\DG}{$^{\circ}$}

\newcommand{\Ed}{$\dot{E}$}

\newcommand{\sig}{$\sigma$}
\newcommand{\Lg}{$L_{\gamma}$}

\newcommand{\psr}{PSR B1259$-$63}

\begin{document}

\title{A Luminous and Highly-variable Gamma-ray Flare Following the 2017 Periastron of PSR B1259$-$63/LS 2883}

\author{T.~J.~Johnson}
\email{tyrel.johnson.ctr@nrl.navy.mil}
\affiliation{College of Science, George Mason University, Fairfax, VA 22030, resident at Naval Research Laboratory, Washington, DC 20375, USA}
\author{K.~S.~Wood}
\email{kent.wood.ctr@nrl.navy.mil}
\affiliation{Praxis Inc., Alexandria, VA 22303, resident at Naval Research Laboratory, Washington, DC 20375, USA}
\author{M.~Kerr}
\affiliation{Space Science Division, Naval Research Laboratory, Washington, DC 20375-5352, USA}
\author{R.~H.~D.~Corbet}
\affiliation{University of Maryland, Baltimore County, and X-ray Astrophysics Laboratory, Code 662 NASA Goddard Space Flight Center, Greenbelt Rd., MD 20771, USA}
\affiliation{Maryland Institute College of Art, 1300 W Mt Royal Ave, Baltimore, MD 21217, USA}
\author{C.~C.~Cheung}
\affiliation{Space Science Division, Naval Research Laboratory, Washington, DC 20375-5352, USA}
\author{P.~S.~Ray}
\affiliation{Space Science Division, Naval Research Laboratory, Washington, DC 20375-5352, USA}
\author{N.~Omodei}
\affiliation{W. W. Hansen Experimental Physics Laboratory, Kavli Institute for Particle Astrophysics and Cosmology, Department of Physics and SLAC National Accelerator Laboratory, Stanford University, Stanford, CA 94305, USA}

\begin{abstract}
Three periastron passages of the PSR B1259$-$63/LS 2883 binary system, consisting of a 48 ms rotation-powered pulsar and a Be star, have been observed by the Large Area Telescope (LAT) on board the \emph{Fermi Gamma-ray Space Telescope}, in 2010, 2014, and 2017.  During the most-recent periastron passage, sustained low-level gamma-ray emission was observed over a $\sim3$-week long interval immediately after periastron, which was followed by an interval of no emission.  Sporadic flares were detected starting 40 days post-periastron and lasted approximately 50 days, during which the emission displayed significant spectral curvature, variability on timescales as short as 1.5 minutes, and peak flux levels well in excess of the pulsar spin-down power.  By contrast, during the 2010 and 2014 periastron passages, significant gamma-ray emission was not observed with the LAT until 30 and 32 days post-periastron, respectively.  The previous flares did not exhibit spectral curvature, showed no short term variability, and did not exceed the pulsar spin-down power.  The high flux and short timescales observed in 2017 suggest significant beaming of the emission is required and constrain the size of the emission region.  The flares occur long enough after periastron that the neutron star should already have passed through the extended disk-like outflow, thus constraining options for target material and seed photon sources for inverse Compton models.
\end{abstract}


\section{INTRODUCTION}\label{intro}
The \psr/LS 2883 \citep{SS71} system consists of a rotation-powered pulsar with a 48 ms spin period \citep[first detected as a radio pulsar by][]{Johnston92} and a Be star in a 3.4 year \citep{Negueruela11}, highly-eccentric orbit \citep[$e = 0.87$,][]{Shannon14}. Pulsations have not been reported at any other wavelength.  The Be star is surrounded by a disk-like outflow \citep[inclined by 10-40\DG\ from the orbit plane,][]{Melatos95} which the pulsar crosses through just before and just after periastron, leading to enhanced emission across the electromagnetic spectrum \citep[e.g.,][]{Chernyakova14,HESS05b}.  No evidence of accretion by the pulsar as it passes through the outflow has ever been reported.

Each time the pulsar crosses through the disk-like outflow of the Be star, the interaction between the pulsar wind and the outflow material leads to emission at radio, X-ray, and TeV wavelengths, producing a double-peaked light curve centered on periastron with the post-periastron hump being more luminous \citep[see Fig.~1 of][for a multi-wavelength light curve]{Chernyakova14}.  The maxima of these peaks occur approximately two weeks before and two weeks after periastron, with the X-rays peaking just before (after) the pre-(post-)periastron radio peak.  In optical, the H$\alpha$ and He I lines, integrated over the disk-like outflow, become increasingly activated and extended as the periastron interactions advance.  The peak X-ray energy flux for the 2010 periastron passage was $\sim4\times10^{-11}$ erg cm$^{-2}$ s$^{-1}$ \citep{Chernyakova14}, while the TeV energy flux was $\sim3\times10^{-12}$ erg cm$^{-2}$ s${-1}$ \citep{HESS05b}.

Launched in June 2008, the \emph{Fermi Gamma-ray Space Telescope} has now been operational during three periastron passages of this system.  The Large Area Telescope \citep[LAT, one of two instruments on board \fermi,][]{LATpaper} has now detected significant enhancements in emission above 0.1 GeV from this system following each periastron passage.  Following the 2010 and 2014 periastron passages, \fermi\ LAT observations established that the GeV portion of the spectrum dominates the spectral energy distribution, with a peak energy flux above 0.1 GeV of $\sim9\times10^{-10}$ erg cm$^{-2}$ s${-1}$, and is also highly variable with weak emission, at best, before periastron and dramatic flaring after \citep{Abdo11,Tam11,Chernyakova14,Tam15,CaliandroB1259}.

The GeV light curve during the 2014  periastron passage was broadly similar to that observed in 2010.  The first detectable emission was observed 30 and 32 days post-periastron, in 2010 and 2014 respectively, rising quickly to a peak flux and then decaying smoothly, with small rises, over the next 40 to 50 days.  In both events, the spectra of the emission were best characterized with power-law shapes falling with energy $E$ approximately as $E^{-3}$.

The 2010 and 2014 events present several challenges to emission models.  Any viable model must be able to produce a gamma-ray luminosity (\Lg) near 100\% of the pulsar spin-down power ($\dot{E} = 8.2\times10^{35}$ erg s$^{-1}$) during the flares \citep[assuming a distance of $d = 2.3$ kpc,][]{Negueruela11}.  Recent interferometric observations of the system have resulted in an increased distance of $2.70^{+0.41}_{-0.31}$ kpc \citep{MJ18}, resulting in \Lg$>$\Ed\ during previous periastron passages.  Models must also explain the lack of a matching GeV flare occurring at approximately the same number of days before periastron.  The absence of corresponding flares at other wavelengths is also a puzzle.

The GeV observations near periastron can be explained using models that invoke either inverse Compton (IC) or Doppler boosted synchrotron emission.  The model of  \citet{Khangulyan12} generates GeV gamma rays via IC interactions between particles in the unshocked pulsar wind, leading to no contemporaneous radio or X-ray flare, and photons from the Be star and/or the disk-like outflow.  The timing of the flare, $\sim$30 days after periastron, is a result of the pulsar wind zone increasing drastically, along the line of sight to Earth, as the pulsar exits the ouflow of the Be star.  Before periastron, the geometry is such that the pulsar wind zone expands in the opposite direction, explaining the lack of a pre-periastron GeV flare.  \citet{Khangulyan12} note that the photon field from the Be star is insufficient to lead to Compotonization of the unshocked pulsar wind quickly enough to account for the near 100\% efficiency of converting spin-down power into gamma rays.  They postulate that the outflow might be heated sufficiently from interactions with the pulsar, before the flare, to sufficiently increase the infrared photon density and match the observed \Lg.

\citet{DC13} also explained the GeV flare as the result of IC interactions, but their model uses electron-positron pairs generated near the pulsar, with a narrow energy distribution, to upscatter X-rays generated in the tail of the shocked pulsar wind.  The X-rays are backscattered towards the Earth after periastron and in the opposite direction before.  Their model does not rely on Be star for target photons and matches the general shape of the GeV flare well, though it fails to produce the time delay between the flares at different wavelengths.

While it is difficult for models invoking IC emission to generate \Lg $\sim$\Ed, Doppler boosted synchrotron emission would naturally explain this result.  \citet{Kong12} proposed that Doppler boosting of X-rays from the tail of the bow shock between the pulsar and Be star winds could explain the GeV flare.  Before periastron, no boosted emission is seen and the synchrotron spectrum should peak below 100 MeV, in agreement with LAT upper limits.  After periastron, a Doppler factor of $\sim$2 produces a spectrum that matched the LAT spectrum well.  This model does not explain the lack of corresponding emission at other wavelengths during the LAT flare nor the timing of the flare $\sim$30 days after periastron, but a possible correlation of the X-ray spectrum with the LAT flare was reported by \citet{Tam15}.  Characterizing the pattern of post-periastron GeV activity, via observing multiple orbital periods, is an important step towards understanding the emission mechanisms and dynamics of this intriguing system.

The most recent periastron passage of the \psr/LS 2883 system occurred on MJD 58018.143 (UTC 2017-09-22 03:25:55.2) and it was quickly apparent that the behavior of the $> 0.1$ GeV emission was different than in 2010 and 2014.  \citet{PrePeriATel} reported a detection using LAT data during the month before periastron.  Detections on daily timescales from 9 to 11 days post-periastron were reported by \citet{FirstATel}.  Following this early emission, no significant gamma-ray flux was detected until 40 days post-periastron.  This renewed activity was more intense and displayed variability on timescales of less than six hours \citep{RobinATel,SecondATel}.  \citet{Tam18} later reported variability on 3 hour time scales.  The emission lasted longer than either of the previous events observed with LAT, including a detection 70 days after periastron with a flux level of $\sim1\times10^{-5}$ cm$^{-2}$ s$^{-1}$ in an eight-hour interval \citep{LastATel}.

We report on LAT observations surrounding the 2017 periastron passage of the \psr/LS 2883 system in more detail.  We confirm the reports of rapid variability, show that the flux changes on timescales as short as $\sim 1.5$ minutes, and report evidence for spectral curvature during the most intense emission.

\section{OBSERVATIONS AND DATA ANALYSIS}\label{obs}
We selected LAT Pass 8 data recorded between MJD 57920.143 and 58129.043 (2017-06-16 03:25:55.2 and 2018-01-11 01:01:55.2 UTC) with reconstructed directions within 15 degrees of (R.A., Dec.; J2000.0) = ($195\fdg699$,$-63\fdg836$) \citep[the radio timing position of \psr,][]{Shannon14}, and energies between 0.1 and 300 GeV.  This time frame provides enough pre-periastron time to obtain an acceptable fit of background sources and lasts until the flux light curve analysis (Section \ref{spec}) gave two weeks of no detections with $\geq 3\sigma$ significance when fitting in one-week and one-day bins.  We kept only those events with measured zenith angles $\leq 90^{\circ}$, to reduce contamination from gamma rays associated with the limb of the Earth, and belonging to the \texttt{SOURCE} class, as defined under the \texttt{P8R2\_SOURCE\_V6} instrument response functions.  Finally, we filtered the data to include only time intervals when the data were marked as good and the observatory was in normal science operations mode.

We created a model of the region including all point and extended sources from the LAT four-year source catalog \citep[3FGL,][]{3FGL} within 25\DG\ of \psr\ and the current Pass 8 Galactic and isotropic diffuse background models\footnote{Available at \url{https://fermi.gsfc.nasa.gov/ssc/data/access/lat/BackgroundModels.html}.} `gll\_iem\_v06.fits' and `iso\_P8R2\_SOURCE\_V6\_v06.txt' \citep{DiffMods}.  \psr\ does not have a 3FGL counterpart, so we added a point source at its position.

\subsection{GAMMA-RAY SPECTRUM AND LIGHT CURVE}\label{spec}
We performed a binned maximum likelihood analysis of a 20\DG$\times$20\DG\ square region centered on \psr, using the spatial and spectral model described previously, over the entire timespan of the dataset.  We allowed the spectral parameters of \psr\ and sources detected in 3FGL with average significance $\geq10\sigma$ and within 6\DG\ of \psr\ to vary.  For those sources flagged as significantly variable in 3FGL and within 8\DG\ of \psr, we allowed the normalization parameters to vary if they didn't pass the previous criteria.  The spectrum of the Galactic diffuse component was modified by a power law, normalization and index free, while the isotropic diffuse component had only a free normalization.  Because the 3FGL catalog was derived from a shorter time span and using less-sensitive Pass 7 reprocessed data, we constructed a 10\DG$\times$10\DG\ test-statistic \citep[TS,][]{Mattox96} map centered on \psr\ to look for unmodeled sources.  No new sources were found in this TS map, suggesting that the use of the 3FGL sources is sufficient.

\begin{figure}[h]
\epsscale{1.0}
\begin{center}
\plotone{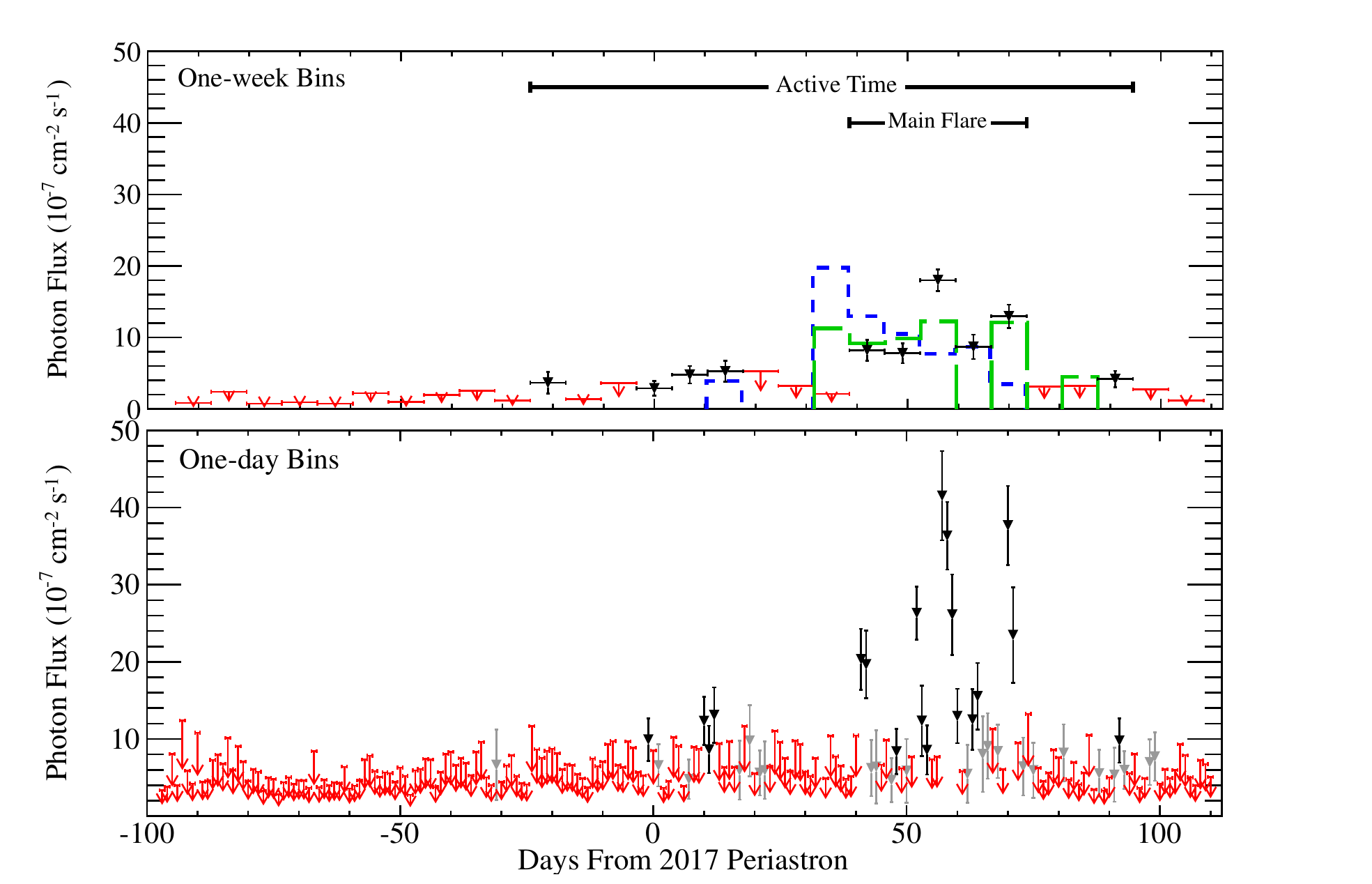}
\end{center}
\caption{Flux light curve of \psr, from 0.1 to 300 GeV, in one-week bins (top) and one-day bins (bottom) covering the 2017 periastron passage.  The vertical axes in both plots have the same dynamic range for better comparison.  Points are plotted for the one-week bins only if \psr\ is found above the background with TS $\geq 12$ and at least 4 predicted counts.  For the one-day bins, we plot black points if \psr\ is found above the background with TS $\geq 9$ and gray points if $4 \leq$ TS $< 9$, both with at least 4 predicted counts.  Otherwise, we show 95\% confidence-level upper limits for both binnings.  In the top panel, the blue histogram with small dashed line shows the 2010 detections and the green histogram with large dashed line shows the 2014 detections, taken from \citet{CaliandroB1259}.  The active time and main flare selections are indicated in the top plot. \label{fig:flxlc}}
\end{figure}

Using the results of our initial fit, we constructed a new model, fixing the parameters of free point and extended sources to the best-fit values but allowing normalizations to vary.  We also fixed the spectral parameters of the Galactic diffuse emission, but left the normalization of the isotropic diffuse component free.  Then, we defined one-week long intervals such that periastron, which occurred on MJD 58018.143 (2017 September 22), was at the center of a bin.  We performed similar maximum likelihood fits in each one-week interval, using the new region model, with the spectrum of \psr\ modeled as a power law, $dN/dE = K (E/E_{0})^{-\Gamma}$, with the normalization $K$ and photon index $\Gamma$ both free to vary while the energy scale value was fixed at $E_{0} = 1$ GeV (Fig.~\ref{fig:flxlc}, top panel).  Our single, pre-periastron detection on one-week timescales corresponds to the time frame when \citet{PrePeriATel} first saw emission rise above TS of 6 in five-day bins.  Our flux light curves agree reasonably well with those of \citet{Tam18}.  We do not see corresponding detections in the two bins just before periastron for our one-week bin light curve, matching their five-day bins, with our TS values being 2 and 6.  This may be due to slight differences in the analysis and our choice to place periastron at the center of a bin as opposed to the start/end of a bin. 

We defined an active time when emission was seen as MJD 57923.643 to 58126.643 and tested for spectral curvature by fitting the spectrum of \psr\ as both a power law and an exponentially cutoff power law, $dN/dE = K (E/E_{0})^{-\Gamma} \exp \lbrace -(E/E_{\rm C} ) \rbrace$, where $E_{\rm C}$ is the cutoff energy.  Following \citet{2PC}, we calculated TS$_{\rm cut}$, testing the significance of the curvature with respect to a power-law shape, and found a preference for the cutoff over the power law with TS$_{\rm cut} = 19.6$ ($> 4\sigma$).  The best-fit spectral values for both are given in Table \ref{tbl:est}, as well as the photon ($F$) and energy ($G$) fluxes integrated from 0.1 to 300 GeV.

\begin{deluxetable}{l c c}
\tablewidth{0pt}
\tablecaption{Best-Fit Spectral Parameters\label{tbl:est}}
\tablecolumns{3}
\tablehead{\colhead{Parameter} & \colhead{Power Law} & \colhead{Cutoff Power Law} }
\startdata
\uline{Active Time:} & & \\
$\Gamma$ & 2.96$\pm$0.06 & 2.41$\pm$0.24 \\
$E_{\rm C}$ (GeV) & \nodata & 0.83$\pm$0.40 \\
$F$ (10$^{-7}$ cm$^{-2}$ s$^{-1}$) & 5.99$\pm$0.35 & 5.86$\pm$0.37 \\
$G$ (10$^{-10}$ erg cm$^{-2}$ s$^{-1}$) & 1.95$\pm$0.12 & 1.91$\pm$0.11 \\
\hline
\uline{Main Flare:} & & \\
$\Gamma$ & 2.78$\pm$0.06 & 2.15$\pm$0.17 \\
$E_{\rm C}$ (GeV) & \nodata & 0.81$\pm$0.24 \\
$F$ (10$^{-7}$ cm$^{-2}$ s$^{-1}$) & 11.7$\pm$0.7 & 11.2$\pm$0.8 \\
$G$ (10$^{-10}$ erg cm$^{-2}$ s$^{-1}$) & 4.26$\pm$0.26 & 4.02$\pm$0.24 \\
\enddata
\tablecomments{Gamma-ray spectral parameters averaged over the active time (top) and main flare (bottom) assuming a power law (column 2) and an exponentially cutoff power law (column 3) for the spectrum of \psr.  The photon $F$ and energy $G$ flux values are integrated over 0.1 to 300 GeV.}
\end{deluxetable}

We then investigated the week with significant emission prior to periastron, the three weeks immediately after periastron, and the main flare from 40 to 75 days after periastron.  During the main flare, a cutoff in the spectrum was significantly preferred over a power-law shape with TS$_{\rm cut} = 26.5$ ($\sim 5\sigma$).  The main flare best-fit parameters are reported in Table \ref{tbl:est}.  The cutoff energy for this interval is consistent with the fit of the entire active time, and the photon index is smaller, though compatible within the uncertainties.  The best-fit cutoff energy and photon index point to a need for observations in the MeV energy range.  In the week before periastron, we found no evidence for curvature in the spectrum (TS$_{\rm cut} = 0$), which is not surprising given that \psr\ is found just at our TS limit of 12 in this interval.  Just after periastron, the total emission is more significant, but we did not find a strong preference for a cutoff with TS$_{\rm cut} = 8.8$, slightly less than 3\sig.

With the expectation that spectral curvature would not be detectable on one-day timescales, we used the best-fit power-law parameters given in Table \ref{tbl:est} to construct a one-day flux light curve.  We only allowed the normalization parameters of \psr, the isotropic diffuse emission, and sources within 3\DG\ of \psr\ to vary.  The resulting flux light curve is shown in the bottom panel of Fig.~\ref{fig:flxlc}.

The pre-periastron detection, on one-week timescales, of Fig.~\ref{fig:flxlc} is temporally consistent with the marginally significant ($5 <$ TS $< 25$) pre-periastron detections, on five-day timescales, found in 2010 and 2014 \citep{Tam15}, though our spectrum is significantly softer.  In the 2010 and 2014 events, the emission started rising earlier than in 2017, reached a maximum over a few days, and then decayed, with a few, lower flux peaks as the emission faded.  As can be seen in the bottom panel of Fig.~\ref{fig:flxlc}, the GeV flare after the 2017 periastron consists of separate peaks, the first of which does not correspond to the highest flux level.  \citet{CaliandroB1259} found maximum one-day flux levels, 36 and 38 days post-periastron, of approximately $30\times10^{-7}$ and $20\times10^{-7}$ cm$^{-2}$ s$^{-1}$ for the 2010 and 2014 events, respectively.  Fig.~\ref{fig:flxlc} shows a maximum one-day flux level of $(42 \pm 6)\times10^{-7}$ cm$^{-2}$ s$^{-1}$ 56 days after periastron.  A more detailed comparison of the gamma-ray and multi-wavelength light curves of these three periastron passages is deferred to a future paper.

\subsection{RAPID VARIABILITY}\label{rv}
\begin{figure}[h]
\epsscale{1.0}
\begin{center}
\plotone{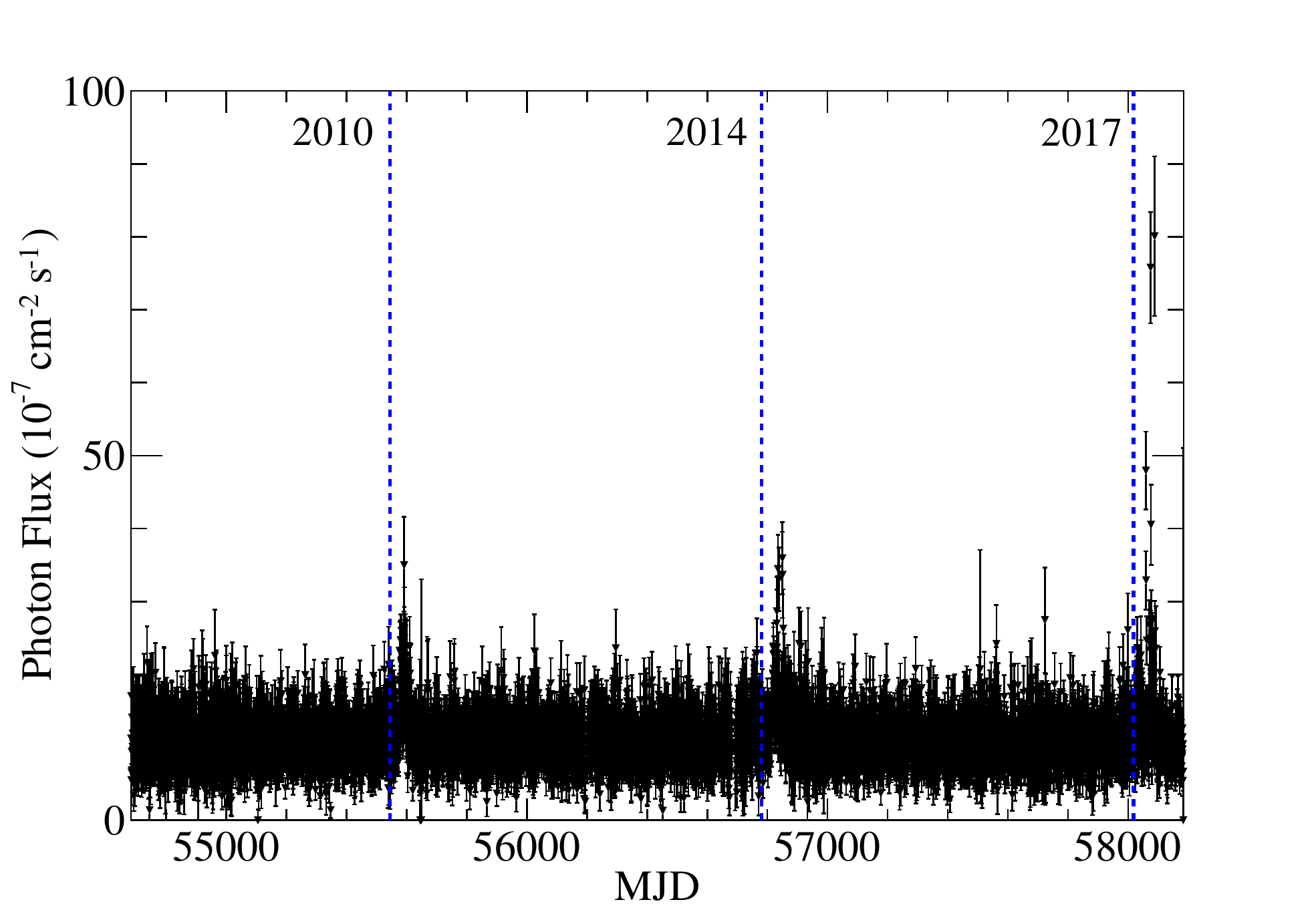}
\end{center}
\caption{Mission-long aperture photometry light curve, in 12 hour bins, of \psr.  The dashed blue vertical lines mark the dates of the three periastrons which have occurred since the launch of \fermi. \label{fig:ap}}
\end{figure}

Following reports of rapid variability \citep{RobinATel,SecondATel}, we performed an aperture photometry analysis spanning nearly the entire \fermi\ mission in 12 hour bins \citep[similar to the analysis in][]{CygX309}.  We used the same data selection criteria described in Section \ref{obs}, but increased the maximum zenith angle to 105\DG\ and restricted events to within 1\DG\ of \psr.  The more strict zenith angle cut described in Section \ref{obs} is necessary for spectral analysis, but tests have shown that periodic signals are stronger with a more lenient selection and any extra contamination in such a small region of interest is acceptable \citep{LMCP3}.  No barycenter correction was applied and no background was subtracted.  The resulting light curve is shown in Fig.~\ref{fig:ap}, all three post-periastron gamma-ray flares are evident.  The aperture photometry suggests that similar, rapid variability was not present during the 2010 and 2014 events, in agreement with the three-hour aperture photometry of \citet{Tam18}.

To characterize any rapid variability, we first created flux light curves on timescales of the \fermi\ spacecraft orbit for each one-day bin with significant emission between 40 and 75 days post-periastron.  We used the spacecraft file to calculate when the zenith angle of \psr\ reached a maximum.  We then defined an `orbit' to be from one maximum to the next.  This definition has the benefit of not breaking up exposure in an orbit, depending on where in the orbital precession period the spacecraft is.

In preparation for unbinned maximum likelihood fits, we included the instrument azimuth angle dependence when calculating the exposure files, as we can not guarantee that the position of the source can be treated as averaging over this coordinate.  To model the region, we used the results from fitting the active time and fixed the parameters of all sources except the normalization and photon index of \psr\ (using a power-law spectral shape), the normalization of the isotropic diffuse emission, and the normalization of 3FGL J1305.7$-$6241, the brightest point source within 3\DG\ of \psr.  We only analyzed data for orbits in which there were at least 10 events recorded within 15\DG\ of \psr.  This last requirement was made to avoid erroneous results giving unrealistically high flux levels, based on experience analyzing the data in these short intervals.

\begin{sidewaysfigure}[h]
\epsscale{1.25}
\begin{center}
\plotone{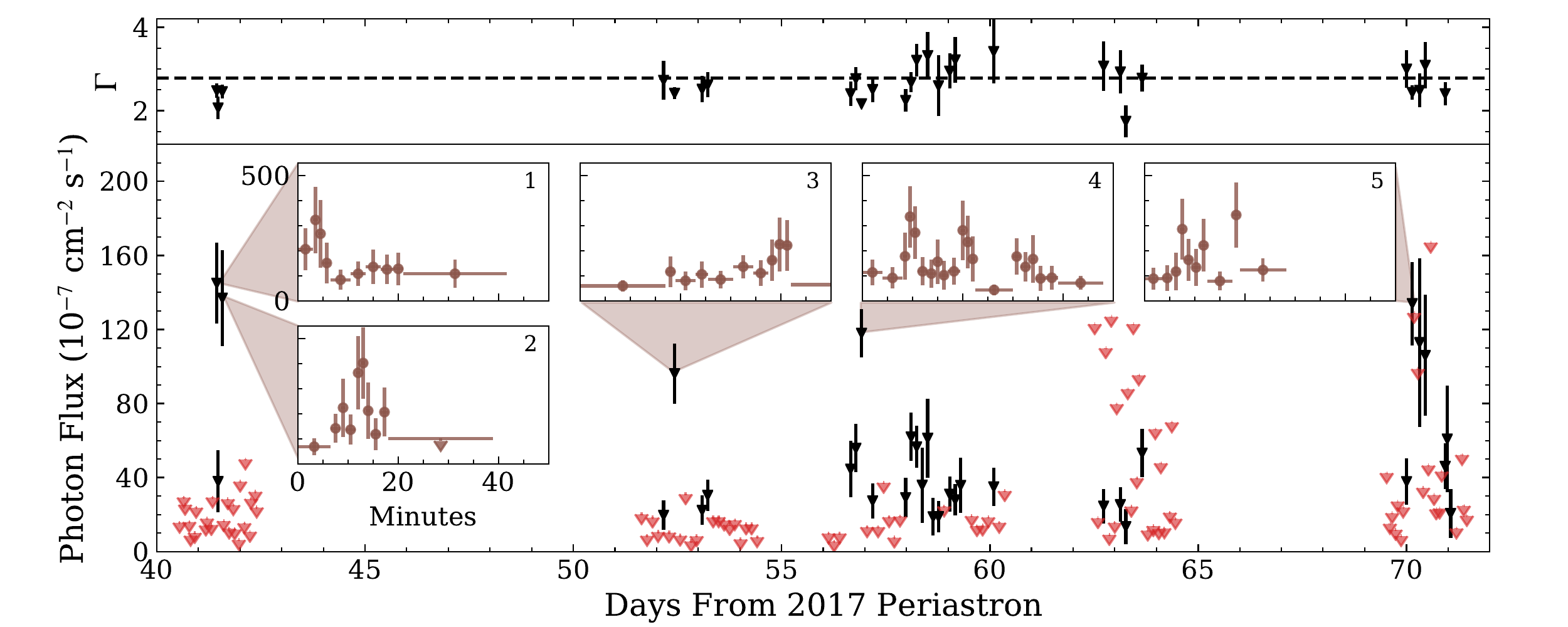}
\end{center}
\caption{Flux light curve of \psr\ in single-orbit time bins, points are plotted only for orbits with TS$\geq 12$, the best-fit photon index is plotted in the topmost panel for those points, else a 95\% confidence-level upper limit is shown.  The insets numbered 1 through 5 show the light curves, requiring TS$\geq 9$ as described in the text, for the five orbits with the highest TS with the vertical axis range matched to that of the main plot and the horizontal axis spanning 50 minutes.  The horizontal dashed line in the top panel shows the best-fit photon index from the main flare period.\label{fig:orblc}}
\end{sidewaysfigure}

The combined orbit-by-orbit flux light curve and photon index variation are shown in Fig.~\ref{fig:orblc}.  Each point is centered on orbit noon, the point in each orbit when \psr\ reaches a minimum zenith angle, and the horizontal extent reflects the time period when \psr\ is within 60\DG\ of the LAT boresight, not the start and stop of the orbit, as this typically gives a better idea of when events could be recorded from the source.

Fig.~\ref{fig:orblc} displays significant variation on timescales as short as one orbit, with flux levels up to $\sim 7$ times what a one-day analysis would suggest.  The photon index stays relatively stable across the orbits we analyzed.

We selected the five orbits with the highest TS values to probe variability on even shorter timescales.  For each of these orbits, we created a 30 s time bin, starting at the time of the first event, and required at least 10 events in the 15\DG\ radius region or else we enlarged the time bin in 30 s steps until this criterion was met.  Next, we performed an unbinned maximum likelihood fit, with only the normalizations of \psr\ and the isotropic diffuse component free to vary, and required that \psr\ be detected with TS$\geq 9$ and at least 4 predicted counts, otherwise we enlarged the bin by 30 s and repeated the fit.  Once those criteria were met, we recorded the best-fit information and started the process anew.  If the last time bin did not meet the TS or predicted counts criteria, we compared the flux of the previous bin to that derived by merging that bin with the last bin.  If these two flux values were comparable, we kept the merged bin, otherwise we report a 95\% confidence-level upper limit for the last bin.  The resulting intra-orbit light curves are shown as numbered insets in Fig.~\ref{fig:orblc}.

For these orbits, the bins range in size from 1 to $\sim 20$ minutes, with flux values doubling or tripling over timescales as small as 1 to 1.5 minutes.  As a measure of the significance of the intra-orbit variability, we calculated $\delta\rm{TS} = \rm{TS}_{\Sigma} - \rm{TS}_{\rm{orb}}$ where TS$_{\rm orb}$ is the TS value found when fitting the entire orbit and TS$_{\Sigma}$ is the sum of the TS values in each intra-orbit bin.  This gave $\Delta\rm{TS} =$ 11, 84, 56, 26, and 61 for insets 1 through 5, in order. Testing against the hypothesis that the flux does not vary over an orbit, the $\Delta\rm{TS}$ values should follow a $\chi^{2}_{\rm{nbin}-1}$ distribution.  Under this assumption we find significances of $1.1\sigma$, $7.6\sigma$, $5.8\sigma$, $1.5\sigma$, and $6.1\sigma$, respectively.  These results suggest that the intra-orbit light curves in insets 1 and 4 show only marginal variability, but this may be due to our method emphasizing the highest time resolution over the most significant binning.

For inset 3, following the prescription described previously resulted in the first bin having an extremely high flux.  However, upon investigation, we discovered that this bin had unusually low exposure and all of the events had zenith angle values $> 80^{\circ}$.  We therefore merged that bin with the one after it, resulting in a more reasonable flux value.  We verified that a similar issue did not occur in the other orbits we analyzed in detail.

\section{DISCUSSION AND CONCLUSIONS}\label{conclsions}
As discussed in Section \ref{intro}, proposed emission models invoke either IC or Doppler boosted synchrotron emission to explain the post-periastron GeV flares from the \psr\ system.  IC models can be hard pressed to generate emission with \Lg$\sim$\Ed, while synchrotron models predict contemporaneous flares at other wavelengths.  The results presented in previous sections provide even more challenges for emission models.

A useful value when discriminating between different emission mechanisms is \Lg/\Ed.  In Table \ref{tbl:energ}, we consider the maximum luminosity on the different time scales we have analyzed, defining \Lg\ = $4\pi G d^{2}$ and assuming a distance of $d = 2.70^{+0.41}_{-0.31}$ kpc.  On daily time scales, the observed gamma-ray luminosity pushes the limits of what the spin-down power, \Ed, can provide.  The variability on even shorter timescales requires significant beaming or an additional source of energy.

\begin{deluxetable}{l c c c}
\tablewidth{0pt}
\tablecaption{Maximum Gamma-ray Energetics on Different Time Scales\label{tbl:energ}}
\tablecolumns{4}
\tablehead{\colhead{Time Scale} & \colhead{$G$} & \colhead{\Lg} & \colhead{\Lg/\Ed}\\
\colhead{} & \colhead{($10^{-10}$ erg cm$^{-2}$ s$^{-1}$)} & \colhead{($10^{35}$ erg s$^{-1}$)} & \colhead{}}
\startdata
One-week & 7.3$\pm$0.6 & 6.4$^{+2.0}_{-1.6}$ & 0.8$\pm$0.2 \\
One-day & 14$\pm$2 & 12$^{+4}_{-3}$  & 1.5$^{+0.5}_{-0.4}$ \\
One-orbit & 70$\pm$16 & 61$^{+18}_{-14}$ & 7.4$^{+2.2}_{-1.7}$ \\
Intra-orbit & 280$\pm$100 & 244$^{+74}_{-56}$ & 29.8$^{+9.0}_{-6.8}$ \\
\enddata
\tablecomments{For the time scales listed during the 2017 periastron passage, this table provides the maximum energy flux ($G$), gamma-ray luminosity (\Lg), and luminosity as a fraction of the spin-down power $\dot{E} = 8.2\times10^{35}$ erg s$^{-1}$ (\Lg/\Ed).  For the uncertainty on \Lg, we incorporate both the energy flux and distance uncertainties.}
\end{deluxetable}

Our results for the 2017 periastron passage disfavor models that generate GeV emission primarily through the IC mechanism, which cannot easily produce \Lg$>$\Ed\ \citep[e.g.,][]{Khangulyan12}.  However, given the stark differences in the progression of the 2017 event from the 2010 and 2014 events, it is possible that a different mechanism was responsible for the emission in 2017, though it is currently unclear why this would be the case.

Following \citet{Tam11}, if we assume that the emission near periastron represents the unboosted synchrotron flux with daily energy flux values of $\sim4\times10^{-10}$ erg cm$^{-2}$ s$^{-1}$ and compare to our maximum intra-orbit energy flux, using Eq.~2 of \citet{Dubus10} the ratio of these fluxes should be $D^{3+\alpha}$, where $D$ is the Doppler factor and $\alpha = \Gamma-1 = 1.17$ using the best-fit photon index from that orbit.  This yields $D\sim3$ as compared to values between 1.5 and 2 found by \citet{Tam11} for the 2010 event.

\citet{Kong12} presented a model for the post-periastron GeV flares as Doppler boosted synchrotron emission and estimated this would result in a cutoff energy of $\sim 0.236\zeta D$ GeV, where $\zeta$ is the electron acceleration efficiency.  They chose a value of $\zeta = 0.36$ so that the synchrotron emission would not contribute significantly above 0.1 GeV before periastron.  If we use the best-fit cutoff energy from the main flare data and $D = 3$, we find $\zeta\sim1.1$, implying a high acceleration efficiency.

Arguably the most exciting result from the 2017 periastron passage is the rapid variability. If we take the shortest rise/fall time of the intra-orbit variability (insets in Fig.~\ref{fig:orblc}) to be the implied light crossing time, $\Delta t \sim 1.5$ minutes, then following \citet{3C279} we can place an upper limit on the radius of the emission region of $R \lesssim D c \Delta t = 8\times10^{7}$ km, where $c$ is the speed of light and using $D = 3$.    Using the orbital inclination angle of $153\fdg4$ derived by \citet{MJ18}, we calculate the distance between the pulsar and the system center of mass, close to the center of the Be star, to be $1.8\times10^{8}$ km and $2.6\times10^{8}$ km 40 and 72 days after periastron, respectively, when we observed intra-orbit variability.  Our upper limit on the radius of the emission region thus varies from $\sim40$\% to $\sim30$\% of the distance between the pulsar and the Be star during the main flare.  New models must therefore explain the flux variation on minute timescales and adhere to this limit on the emission region.

We have shown that the GeV gamma-ray emission associated with the 2017 periastron passage of the \psr\ system progressed in a wholly different manner than the 2010 and 2014 events.  Significant emission was seen for $\sim3$ weeks immediately following periastron and then not again until 40 days after periastron when several flares, lasting a few days each and reaching daily integral photon fluxes as high as $4\times10^{-6}$ cm$^{-2}$ s$^{-1}$, were observed.  The flares demonstrated significant spectral curvature and variability on timescales shorter than the \fermi\ spacecraft orbital period with integral photon flux levels, on minute timescales, as high as $\sim4 \times10^{-5}$ cm$^{-2}$ s$^{-1}$.  Excluding gamma-ray bursts and solar flares, this is the fastest GeV variability observed in LAT data.  Our results challenge the existing emission models of the \psr/LS 2883 system, during the periastron passage and the few months after, and demonstrate the need for continued monitoring over additional orbital cycles.

\acknowledgments

\begin{center}
\emph{ACKNOWLEDGMENTS}
\end{center}

The \fermi\ LAT Collaboration acknowledges generous ongoing support from a number of agencies and institutes that have supported both the development and the operation of the LAT as well as scientific data analysis.  These include the National Aeronautics and Space Administration and the Department of Energy in the United States, the Commissariat \`a l'Energie Atomique and the Centre National de la Recherche Scientifique / Institut National de Physique Nucl\'eaire et de Physique des Particules in France, the Agenzia Spaziale Italiana and the Istituto Nazionale di Fisica Nucleare in Italy, the Ministry of Education, Culture, Sports, Science and Technology (MEXT), High Energy Accelerator Research Organization (KEK) and Japan Aerospace Exploration Agency (JAXA) in Japan, and the K.~A.~Wallenberg Foundation, the Swedish Research Council and the Swedish National Space Board in Sweden.
 
Additional support for science analysis during the operations phase is gratefully acknowledged from the Istituto Nazionale di Astrofisica in Italy and the Centre National d'\'Etudes Spatiales in France. This work performed in part under DOE Contract DE-AC02-76SF00515.

Portions of this research performed at the Naval Research Laboratory are sponsored by NASA DPR S-15633-Y and \fermi\ Guest Investigator grant 16-Fermi10-0006.

\facility{Fermi(LAT)}

\bibliographystyle{aasjournal}
\bibliography{references}

\end{document}